\documentclass[utf8]{frontiersFPHY} 

\usepackage{isotope}
\usepackage{url,hyperref,lineno,microtype,subcaption}
\usepackage[onehalfspacing]{setspace}
\usepackage{amsmath}
\usepackage{changes}

\renewcommand{\sout}{\bgroup \color{red} \ULdepth=-.5ex \ULset}

\allowdisplaybreaks[4]

\def\keyFont{\fontsize{8}{11}\helveticabold }
\def\firstAuthorLast{Rui Wang {et~al.}} 
\def\Authors{Rui Wang\,$^{1,2}$, Zhen Zhang\,$^{3}$, Lie-Wen Chen\,$^{4,*}$, and Yu-Gang Ma\,$^{1,2}$}
\def\Address{$^{1}$Key Laboratory of Nuclear Physics and Ion-beam Application~(MOE), Institute of Modern Physics, Fudan University, Shanghai $200433$, China  \\
$^{2}$Shanghai Institute of Applied Physics, Chinese Academy of Sciences, Shanghai $201800$, China  \\
$^{3}$Sino-French Institute of Nuclear Engineering and Technology, Sun Yat-Sen University, Zhuhai $519082$, China \\
$^{4}$School of Physics and Astronomy and Shanghai Key Laboratory for
Particle Physics and Cosmology, Shanghai Jiao Tong University, Shanghai $200240$, China
}
\def\corrAuthor{Lie-Wen Chen}

\def\corrEmail{lwchen@sjtu.edu.cn}

\begin{document}
\onecolumn
\firstpage{1}

\title[Nuclear collective dynamics in transport model with the lattice Hamiltonian method]{Nuclear collective dynamics in transport model with the lattice Hamiltonian method}

\author[\firstAuthorLast ]{\Authors} 
\address{} 
\correspondance{} 

\extraAuth{}

\maketitle

\begin{abstract}

\section{}
We review the recent progress on studying the nuclear collective dynamics by solving the Boltzmann-Uehling-Uhlenbeck~(BUU) equation with the lattice Hamiltonian method treating the collision term by the full-ensemble stochastic collision approach.
This lattice BUU (LBUU) method has recently been developed and implemented in a GPU parallel computing technique, and achieves a rather stable nuclear ground-state evolution and high accuracy in evaluating the nucleon-nucleon (NN) collision term.
This new LBUU method has been applied to investigate the nuclear isoscalar giant monopole resonances and isovector giant dipole resonances.
While the calculations with the LBUU method without the NN collision term (i.e., the lattice Hamiltonian Vlasov method) describe reasonably the excitation energies of nuclear giant resonances, the full LBUU calculations can well reproduce the width of the giant dipole resonance of $^{208}$Pb by including a collisional damping from NN scattering.
The observed strong correlation between the width of nuclear giant dipole resonance and the NN elastic cross section suggests that the NN elastic scattering plays an important role in nuclear collective dynamics, and the width of nuclear giant dipole resonance provides a good probe of the in-medium NN elastic cross section.

\tiny
 \keyFont{ \section{Keywords:} Boltzmann-Uehling-Uhlenbeck equation, Lattice Hamiltonian method, Nuclear giant resonances, Thomas-Fermi initialization, Stochastic collision approach} 
\end{abstract}

\section{Introduction}

Transport models deal with the time-evolution of the Wigner function or phase-space distribution function $f(\vec{r},\vec{p},t)$ which stems from the Wigner representation of the Schr$\ddot{\rm o}$dinger equation~\cite{CarRMP55,BerPR160}, and provide a successful semi-classical time-dependent approach to nuclear dynamics, especially to heavy-ion collisions~(HICs).
One of the main ingredients of transport models is the mean field potential which embodies information on the nuclear equation of state~(EOS) or the in-medium effective nuclear interaction.
Therefore, transport models serve as an important theoretical tool to explore the EOS of asymmetric nuclear matter from observables in HICs.
A lot of information on the nuclear EOS from sub-saturation~\cite{LBAPRL85,CLWPRL94,TsaPRL102} to supra-saturation densities up to about $3\sim5$ times saturation density~\cite{DanPRL81,LBAPRL88,Dansc298,XZPRL102,FZQPLB683,RusPLB697,XWJPLB718,CozPRC88,RusPRC94,CozEPJA54,FenNST29,LiNST29,WolUnv4},
has been obtained from transport model analyses of various obseverables, e.g., collective flows and particle production, in intermediate and high energy HICs.
The exact information about the nuclear EOS is crucial in describing reaction dynamics of exotic nuclei~\cite{BarPR410,LBAPR464}, various properties of both finite nuclei (e.g., neutron skin thickness~\cite{BroPRL85,TypPRC64,YosPRC69} and drip lines~\cite{OyaPRC82,WRPRC92}) and neutron stars (e.g., masses and cooling mechanisms~\cite{PraPRL61,LatPRL66,Latsc304,StePR411,LatPR442}), and astrophysical processes such as supernova explosion scenarios~\cite{SumApJ422,OerRMP89,PaiNST29}.
In particular, it should be mentioned that the first gravitational wave signal GW170817~\cite{LIGO17} of binary neutron star merger has been recently observed and localized by the LIGO and Virgo observatories, and it inaugurates a new era of multimessenger astronomy and gives important constraints on the dense nuclear matter EOS~\cite{ZhouEP18,DePRL121,ZhaNST29,BLEPJA55,ZhouYPRD99}. Moreover, very recently, using the X-ray data from NASA's Neutron Star Interior Composition Explorer (NICER), the mass and radius of the millisecond pulsar PSR J0030+0451 have been simultaneously estimated~\cite{RilApJL19,MilApJL19} and its implications on the dense nuclear matter EOS has been analyzed~\cite{RaaApJL19}. In addition, a new record for the maximum mass of neutron stars, namely, a millisecond pulsar J0740+6620 with mass $2.14^{+0.10}_{-0.09}M_\odot$ ($68.3\%$ credibility interval), has been recently reported~\cite{Cro19NA}, and this heaviest neutron star observed so far can rule out many soft nuclear matter EOS's and especially the supersoft high-density symmetry energy~\cite{ZhouYApJL19}.

The time-dependent-Hartree-Fock~(TDHF) theory provides a very successful quantum many-body approach at the mean-field level to describe low-energy nuclear reaction dynamics including the nuclear collective dynamics~(see, e.g., Refs.~\cite{SimPPNP18,StePPNP19} for recent review).
Given that the Vlasov equation, i.e., Boltzmann-Uehling-Uhlenbeck~(BUU) equation without the nucleon-nucleon (NN) collision term, corresponds to the semi-classical limit of the TDHF equation, transport models can thus be seen as an efficient semi-classical approach to study nuclear collective dynamics.
In particular, the two-particle-two-hole ($2p$-$2h$) correlation beyond the mean-field approximation, which dominates the collisional damping of nuclear giant resonances, can be effectively taken into account in transport models via binary collisions.
In literature, there have been a lot of works studying nuclear giant resonances based on the pure Vlasov equation~\cite{UrbPRC85,BarEPJD68,ZHPRC94}, the Vlasov equation with a collision relaxation time~\cite{SmePRC44}, and the full transport model with both the mean-field and the NN scatterings~\cite{GaiPRC81,TCPRC88,WKPRC95}.
For example, based on simulations of transport models, the excitation energies of nuclear giant resonances have been used to extract information on the nuclear EOS and neutron-proton effective mass splitting~\cite{KHYPRC95}, while the width of nuclear giant dipole resonances~(GDR) has been proposed as an effective probe of the in-medium NN elastic cross section~\cite{WRX19}.
The width of nuclear GDR can also serve as a fingerprint of $\alpha$-particle clustering configurations in nuclei~\cite{HWBPRL113}.

Although transport models have been extensively used in the study of nuclear giant resonances, the accurate description of the giant resonances within transport models is still a challenge.
In transport model calculations, unlike the simulations of HICs at intermediate and high energies, the calculation of nuclear giant resonances, which are the collective excitation states with an excitation energy of about $20$~MeV, requires a more proper description of nuclear ground-state and a rather accurate implementation of Pauli blocking.
In particular, the Pauli blocking is intimately related to the collisional damping and thus the width of nuclear giant resonances in the transport model calculations.
In this sense, the nuclear collective motion provides an ideal topic to examine and improve transport models, since the effects of several deficiencies, such as the inaccurate treatment of Pauli blocking, are more pronounced in nuclear collective dynamics with small amplitude oscillations.
Transport models for HICs can be roughly divided into two categories, the BUU equation~(see, e.g., Ref.~\cite{BerPR160}) and the quantum molecular dynamics~(QMD) model (see, e.g., Ref.~\cite{AicPR202}).
From the viewpoint of transport models, the essential difference between these two types is that the BUU-type transport models mimic $f(\vec{r},\vec{p},t)$ by a large number of ensembles or test particles for each nucleon while the QMD-type by a Gaussian wave packet for each nucleon.
Recently, the transport model community has started the code comparison project~\cite{XJPRC93,ZYXPRC97,OnoPRC100} to try to understand the source of the discrepancies in various transport model codes, and eventually reduce the uncertainties of transport models.
For the issue of Pauli blocking, the QMD-type transport models seem not to be as good as BUU-type models~\cite{ZYXPRC97}, therefore the BUU-type transport models are more suitable for the study of nuclear collective motions, especially for the calculation of the spreading width, in which the accurate treatment of Pauli blocking is essential.

In order to study (near-)equilibrium nuclear dynamics within the framework of transport models, a BUU-type transport model, namely, the lattice BUU~(LBUU) method~\cite{WRPRC99,WRX19}, has been recently developed, and it can achieve good stability for the ground state evolution~\cite{WRPRC99} and treat Pauli blocking with very high accuracy~\cite{WRX19}.
The resulting LBUU framework has the following features:
1) a smearing of the local density, which is common practice in transport models to obtain a smooth mean-field, is included self-consistently in the equations of motion through the lattice Hamiltonian (LH) method;
2) the ground state of a nucleus is obtained through varying the total energy with respect to the nucleon density distribution based on the same Hamiltonian that governs the system evolution;
3) the NN collision term in BUU equation is implemented through a full-ensemble stochastic collision approach.
The above techniques as well as the sufficiently large number of ensembles make it possible to solve the BUU equation almost exactly, and thus one can obtain very accurate results for the nuclear collective motions within the BUU equation.
We note that the high accuracy of the LBUU method relies on increasing computational resources, therefore the high-performance GPU parallel computing~\cite{Rue2013} has been employed in the LBUU framework to improve the computing efficiency.

This paper is organized as follows.
In Sec.~\ref{S:th}, we first introduce the LBUU method for solving the BUU equation, including the mean-field, the collision integral and the initialization for the nuclear ground state, and then describe how to deal with the nuclear giant resonances within transport models.
In Sec.~\ref{S:LHV}, we present results on the peak energies of the nuclear giant resonances from lattice Hamiltonian Vlasov (LHV) calculations, i.e., the LBUU calculations without the NN collision term, and then compare these with the results from the random-phase approximation~(RPA).
In Sec.~\ref{S:width}, we show the results of the strength function and the width of GDR from the full LBUU calculations, and compare these to the experimental data from $\isotope[208]{Pb}(\vec{p},\vec{p}')$ reaction carried out at the Research Center for Nuclear Physics in Osaka, Japan~(RCNP)~\cite{TamPRL107}.
Finally we give a brief summary and outlook in Sec.~\ref{S:S&O}.

\section{Model description}\label{S:th}

The starting point for calculating the nuclear collective motion is the BUU equation with a momentum-dependent mean-field potential $U(\vec{r},\vec{p})$, i.e.,
\begin{equation}\label{E:BUUE}
    \frac{\partial f}{\partial t} + \frac{\vec{p}}{E}\cdot\nabla_{\vec{r}}f + \nabla_{\vec{p}}U(\vec{r},\vec{p})\cdot\nabla_{\vec{r}}f - \nabla_{\vec{r}}U(\vec{r},\vec{p})\cdot\nabla_{\vec{p}}f = I_{\rm c},
\end{equation}
where $f$~(i.e., Wigner function) is the Fourier transform of one-body density matrix $\rho(\vec{r}+\vec{s}/2,\vec{r}-\vec{s}/2)$, i.e.,
\begin{equation}
    f(\vec{r},\vec{p}) = \frac{1}{(2\pi\hbar)^3}\int {\rm exp}\Big(-i\frac{\vec{p}}{\hbar}\cdot\vec{s}\Big)\rho(\vec{r}+\vec{s}/2,\vec{r}-\vec{s}/2)d^3s.
\end{equation}
In the local density approximation, $f(\vec{r},\vec{p})$ is reduced to the classical one-body phase-space distribution function.
The collision term $I_{\rm c}$, which takes into account the Pauli principle due to nucleons' Fermi statistics, reads
\begin{equation}
\begin{split}
    I_{\rm c} = &~-g\int\frac{d^2p_2}{(2\pi\hbar)^3}\frac{d^3p_3}{(2\pi\hbar)^3}\frac{d^3p_4}{(2\pi\hbar)^3}|\mathcal{M}_{12\rightarrow34}|^2(2\pi)^4\delta^4(p_1 + p_2 - p_3 - p_4)\\
    &\times[f_1f_2(1-f_3)(1-f_4) - f_3f_4(1 - f_1)(1 - f_2)],
\end{split}\label{E:Ic}
\end{equation}
where $g$ $=$ $2$ is the spin degeneracy factor, and $\mathcal{M}_{12\rightarrow34}$ is the in-medium transition matrix element.
Note that we have ignored the isospin index in the above three equations, but it can be restored easily.
The BUU equation without the collision term $I_{\rm c}$ is referred to as the Vlasov equation, which is the semi-classical limit of the quantum transport theory with the system described by the one-body phase-space distribution function~\cite{CarRMP55,BerPR160}, whereas the quantum corrections can be included perturbatively~\cite{BonPRL71,KonNPA577}.

We use the LH method, originally proposed by Lenk and Pandharipande~\cite{LenPRC39} in 1989, to solve the BUU equation.
The LH method has been successfully employed in the study of HICs~\cite{XHMPRL65,XHMPRL67}.
It improves the sample smoothing technique of the usual test particle approach~\cite{WonPRC25}, and conserves the total energy almost exactly.
In the LH method, the phase-space distribution function $f_{\tau}(\vec{r},\vec{p},t)$ is mimicked by $A\times N_{\rm E}$ test nucleons with a form factor $S$ in the coordinate space to modify the relation between the test nucleons and the Wigner function, i.e.,
\begin{equation}\label{E:f}
    f_{\tau}(\vec{r},\vec{p},t) = \frac{1}{g}\frac{(2\pi\hbar)^3}{N_{\rm E}}\sum_i^{AN_{\rm E},\tau}S\big[\vec{r}_i(t) - \vec{r}\big]\delta\big[\vec{p}_i(t) - \vec{p}\big],
\end{equation}
where $A$ is the mass number of the system, and $N_{\rm E}$ is the number of ensembles or number of test particles, usually a very large number, used in the calculation.
The sum in the above expression runs over all test nucleons with isospin $\tau$.
The form factor $S$ can take a Gaussian form, or a certain form with a finite range that ensures the particle number conservation.
By giving each test nucleon a form factor, the movement of a test nucleon leads to a continuous variation of the local nucleon density of the nearby lattice sites, which is useful to smoothen the nucleon distribution functions in phase space.
A similar form factor in momentum space [here the $\delta$-function is used in Eq.~(\ref{E:f})] could be introduced and this might be helpful to reduce fluctuations if momentum-dependent mean-field potential is employed, and it would be interesting to carry out a systematic investigation of the effects of a form factor in momentum space in the future.
The equations of motion of the test nucleons are governed by the total Hamiltonian, and we approximate the latter by the lattice Hamiltonian, i.e.,
\begin{equation}\label{E:HL}
    H = \int {\mathcal{H}}(\vec{r})d\vec{r} \approx l_xl_yl_z\sum_{\alpha}{\mathcal{H}}(\vec{r}_{\alpha})\equiv H_L,
\end{equation}
where $\vec{r}_{\alpha}$ denotes the coordinate of lattice site $\alpha$, and $l_x$, $l_y$ and $l_z$ are the lattice spacings.
Therefore in the LH method, only the values of the phase-space distribution function at lattice sites $f_{\tau}(\vec{r}_{\alpha},\vec{p},t)$ need to be calculated.

By solving the BUU equation or Vlasov equation based on the LH method, one obtains the time evolution of $f(\vec{r},\vec{p},t)$, or the test nucleons' coordinates $\vec{r}_i$ and momenta $\vec{p}_i$, and then the time evolution of other physical quantities can be calculated accordingly.

\subsection{Mean fields}

We employ the Skyrme pseudopotential to calculate the lattice Hamiltonian in Eq.~(\ref{E:HL}).
The next-to-next-to-next leading order~(N$3$LO) Skyrme pseudopotential~\cite{RaiPRC83}, which is a mapping of N$3$LO local energy density functional~\cite{CarPRC78}, generalizes the standard Skyrme interaction~\cite{ChaNPA627} and can reproduce the empirical nuclear optical potential up to about $1$~GeV in kinetic energy~\cite{WRPRC98}, for which the standard Skyrme interactions fail to describe.
The Hamiltonian density from the N$3$LO Skyrme pseudopotential contains the kinetic term $\mathcal{H}^{\rm kin}(\vec{r})$, the local term $\mathcal{H}^{\rm loc}(\vec{r})$, the momentum-dependent term $\mathcal{H}^{\rm MD}(\vec{r})$, the density-dependent term $\mathcal{H}^{\rm DD}(\vec{r})$ and the gradient term $\mathcal{H}^{\rm grad}(\vec{r})$.
The kinetic term
\begin{equation}\label{E:Hkin}
    {\mathcal H}^{\rm kin}(\vec{r}) = \sum_{\tau = n,p}\int d^3p\frac{p^2}{2m_{\tau}}f_{\tau}(\vec{r},\vec{p}),
\end{equation}
and the local term
\begin{equation}\label{E:Hloc}
    {\mathcal H}^{\rm loc}(\vec{r}) = \frac{t_0}{4}\bigg[(2 + x_0)\rho^2 - (2x_0 + 1)\sum_{\tau = n,p}\rho_{\tau}^2\bigg],
\end{equation}
are the same as those from the standard Skyrme interaction.
The momentum-dependent term is written in the following form,
\begin{equation}\label{E:HMD}
    {\mathcal H}^{\rm MD}(\vec{r}) = \int d^3pd^3p'{\mathcal K}_s(\vec{p},\vec{p}')f(\vec{r},\vec{p})f(\vec{r},\vec{p}') + \sum_{\tau = n, p}\int d^3pd^3p'{\mathcal K}_v(\vec{p},\vec{p}')f_{\tau}(\vec{r},\vec{p})f_{\tau}(\vec{r},\vec{p}'),
\end{equation}
with $f(\vec{r},\vec{p})$ $=$ $f_n(\vec{r},\vec{p})$ $+$ $f_p(\vec{r},\vec{p})$.
The quantities ${\mathcal K}_{\rm s}(\vec{p},\vec{p}')$ and ${\mathcal K}_{\rm v}(\vec{p},\vec{p}')$ in Eq.~(\ref{E:HMD}) represent the isoscalar and isovector kernels of the momentum-dependent part of the mean-field potential, respectively.
${\mathcal K}_{\rm s}(\vec{p},\vec{p}')$ and ${\mathcal K}_{\rm v}(\vec{p},\vec{p}')$ for the N$3$LO Skyrme pseudopotential are expressed as
\begin{align}
    {\mathcal K}_{\rm s}(\vec{p},\vec{p}') & = \frac{C^{[2]}}{16\hbar^2}(\vec{p} - \vec{p}')^2 + \frac{C^{[4]}}{32\hbar^2}(\vec{p} - \vec{p}')^4 + \frac{C^{[6]}}{16\hbar^2}(\vec{p} - \vec{p}')^6,\label{E:mdks}\\
    {\mathcal K}_{\rm v}(\vec{p},\vec{p}') & = \frac{D^{[2]}}{16\hbar^2}(\vec{p} - \vec{p}')^2 + \frac{D^{[4]}}{32\hbar^2}(\vec{p} - \vec{p}')^4 + \frac{D^{[6]}}{16\hbar^2}(\vec{p} - \vec{p}')^6.\label{E:mdkv}
\end{align}
If we keep only the $C^{[2]}$ and $D^{[2]}$ terms, the N$3$LO Skyrme pseudopotential reduces to the standard Skyrme effective interaction.
For the sake of simplicity of performing numerical derivatives, we truncate at the second order of the spatial gradient of $\rho(\vec{r})$,
\begin{align}\label{E:Hgrd}
{\mathcal H}^{\rm grad}(\vec{r}) = &~\frac{1}{8}E^{[2]}\Big\{\rho(\vec{r})\nabla^2\rho(\vec{r}) - \big[\nabla\rho(\vec{r})\big]^2\Big\} + \frac{1}{8}F^{[2]}\sum_{\tau = n,p}\Big\{\rho_{\tau}(\vec{r})\nabla^2\rho_{\tau}(\vec{r}) - \big[\nabla\rho_{\tau}(\vec{r})\big]^2\Big\}\notag\\
 = &~\frac{1}{8}g^{[2]}\Big\{\rho(\vec{r})\nabla^2\rho(\vec{r}) - \big[\nabla\rho(\vec{r})\big]^2\Big\} + \frac{1}{8}g_{\rm iso}^{[2]}\Big\{\rho_{\delta}(\vec{r})\nabla^2\rho_{\delta}(\vec{r}) - \big[\nabla\rho_{\delta}(\vec{r})\big]^2\Big\}.
\end{align}
In the second line we have introduced, $g^{[2]}$ $=$ $E^{[2]} + \frac{1}{2}F^{[2]}$, $g_{\rm iso}^{[2]}$ $=$ $\frac{1}{2}F^{[2]}$, and $\rho_{\delta}$ $=$ $\rho_n-\rho_p$.
We neglect the second term in Eq.~(\ref{E:Hgrd}) since it is much smaller than the first term, in other words, we keep only the second order spatial derivative of the total nucleon density $\rho(\vec{r})$.
The density-dependent term for the N$3$LO Skyrme pseudopotential takes its form in the standard Skyrme interaction
\begin{equation}\label{E:HDD}
\begin{split}
    {\mathcal H}^{\rm DD}(\vec{r}) = \frac{t_3}{24}\bigg[(2 + x_3)\rho^2 - (2x_3 + 1)\sum_{\tau=n,p}\rho_{\tau}^2\bigg]\rho^{\alpha}.
\end{split}
\end{equation}
One can see that the Hamiltonian density ${\mathcal H}(\vec{r})$, expressed as the sum of Eqs. (\ref{E:Hkin})-(\ref{E:HMD}), Eqs. (\ref{E:Hgrd}) and (\ref{E:HDD}), is explicitly dependent on $f_{\tau}(\vec{r},\vec{p})$, as well as the densities $\rho_{\tau}(\vec{r})$ and their derivatives.

In the above expressions, the parameters $C^{[n]}$, $D^{[n]}$, $E^{[n]}$ and $F^{[n]}$ are recombinations of the Skyrme parameters $t_1^{[n]}$, $t_2^{[n]}$, $x_1^{[n]}$ and $x_2^{[n]}$, which are related to derivative terms in the Skyrme two-body potential $v^{\rm Sk}(\vec{r}_1,\vec{r}_2)$~\cite{ChaNPA627}, i.e.,
\begin{align}
C^{[n]} & = t_1^{[n]}(2+x_1^{[n]})+t_2^{[n]}(2+x_2^{[n]}),\label{E:Cn}\\
D^{[n]} & = -t_1^{[n]}(2x_1^{[n]}+1)+t_2^{[n]}(2x_2^{[n]}+1),\label{E:Dn}\\
E^{[n]} & = \frac{i^n}{2^n}\big[t_1^{[n]}(2+x_1^{[n]}) - t_2^{[n]}(2+x_2^{[n]})\Big],\label{E:En}\\
F^{[n]} & = -\frac{i^n}{2^n}\big[t_1^{[n]}(2x_1^{[n]}+1) + t_2^{[n]}(2x_2^{[n]}+1)\Big]\label{E:Fn}.
\end{align}
Specifically, we obtain the coefficient of the gradient term
\begin{equation}
    g^{[2]} = E^{[2]} + \frac{1}{2}F^{[2]} = -\frac{1}{8}\Big[3t_1^{[2]} - t_2^{[2]}(5 + 4x_2^{[2]})\Big].\label{E:g2}
\end{equation}

Substituting $f(\vec{r},\vec{p},t)$ as expressed in Eq.~(\ref{E:f}) into Eqs.~(\ref{E:Hkin})-(\ref{E:HDD}), and noting that the local nucleon density $\rho_{\tau}(\vec{r})$ is given by integrating $f_{\tau}(\vec{r},\vec{p},t)$ with respect to momentum,
\begin{equation}\label{E:rhoL}
    \rho_{\tau}(\vec{r},t) = g\int f_{\tau}(\vec{r},\vec{p},t)\frac{d^3p}{(2\pi\hbar)^3} = \frac{1}{N_{\rm E}}\sum_i^{\alpha,\tau}S\big[\vec{r}_i(t) - \vec{r}\big],
\end{equation}
we can express the lattice Hamiltonian $H_L$ in Eq.~(\ref{E:HL}) in terms of the coordinates and momenta of the test nucleons.
Since the coordinate and momentum of the test nucleons $\vec{r}_i$ and $\vec{p}_i$ can be regarded as the canonical variables of the lattice Hamiltonian, their time evolution is then governed by the Hamilton equation for all ensembles,
\begin{align}
\frac{d\vec{r}_i}{dt} = &~N_{\rm E}\frac{\partial H_L\big[\vec{r}_1(t),\cdots,\vec{r}_{A\times N_{\rm E}}(t);\vec{p}_1(t),\cdots,\vec{p}_{A\times N_{\rm E}}(t)\big]}{\partial\vec{p}_i} = \frac{\vec{p}_i(t)}{m} + N_{\rm E}l_xl_yl_z\sum_{\alpha\in V_i}\frac{\partial{\mathcal{H}}^{\rm MD}_{\alpha}}{\partial\vec{p}_i}\label{E:ri},\\
\frac{d\vec{p}_i}{dt} = &~- N_{\rm E}\frac{\partial H_L\big[\vec{r}_1(t),\cdots,\vec{r}_{A\times N_{\rm E}}(t);\vec{p}_1(t),\cdots,\vec{p}_{A\times N_{\rm E}}(t)\big]}{\partial\vec{r}_i}\notag\\
= &~- N_{\rm E}l_xl_yl_z\times\notag\\
&~\sum_{\alpha\in V_i}\bigg\{\sum_{\tau}^{n,p}\bigg[\frac{\partial(\mathcal{H}^{\rm loc}_{\alpha} + \mathcal{H}^{\rm Cou}_{\alpha} + \mathcal{H}^{\rm DD}_{\alpha})}{\partial\rho_{\tau,\alpha}} + \sum_{n = 0}(-1)^n\nabla^n\frac{\partial{\mathcal{H}}^{\rm grad}_{\alpha}}{\partial\nabla^n\rho_{\tau,\alpha}} \bigg]\frac{\partial\rho_{\tau,\alpha}}{\partial\vec{r}_i} + \frac{\partial{\mathcal{H}}^{\rm MD}_{\alpha}}{\partial\vec{r}_i}\bigg\}\label{E:pi}.
\end{align}

In the above two equations, the quantities with a subscript $\alpha$ refers to their values at lattice site $\alpha$.
The $V_i$ under the summation represents the volume, which the form factor of the $i$-th test nucleon covers, and the sums run over all lattice sites inside $V_i$.
The Coulomb interaction contributes to the Hamiltonian density through the term
\begin{eqnarray}\label{E:Cou}
     {\mathcal{H}}^{\rm Cou}(\vec{r}_{\alpha}) &=& e^2\rho_p(\vec{r}_{\alpha})\bigg\{\frac{1}{2}\int\frac{\rho_p(\vec{r}')}{|\vec{r}_{\alpha} - \vec{r}'|}d\vec{r}' - \frac{3}{4}\Big[\frac{3\rho_p(\vec{r}_{\alpha})}{\pi}\Big]^{1/3}\bigg\} \notag\\
    &\approx& e^2\rho_p(\vec{r}_{\alpha})\bigg\{\frac{1}{2}\sum_{\alpha'\ne\alpha}\frac{\rho_p(\vec{r}_{\alpha'})l_xl_yl_z}{|\vec{r}_{\alpha} - \vec{r}_{\alpha'}|} - \frac{3}{4}\Big[\frac{3\rho_p(\vec{r}_{\alpha})}{\pi}\Big]^{1/3}\bigg\},
\end{eqnarray}
where the second term represents the contribution from the Coulomb exchange energy.
Further tests show that the Coulomb energy $\mathcal{H}^{\rm Cou}(\vec{r}_{\alpha})$ converges at a lattice spacing of $l_x$ $=$ $l_y$ $=$ $l_z$ $=$ $0.5~\rm fm$ used in the present LBUU simulations.
The gradient term $\mathcal{H}_{\alpha}^{\rm grad}$ in Eq.~(\ref{E:pi}) is obtained by considering
\begin{align}
    \delta\int\mathcal{H}^{\rm grad}(\vec{r})d^3r = &~\sum_{\tau}^{n,p}\int\bigg[\frac{\partial\mathcal{H}^{\rm grad}(\vec{r})}{\partial\rho_{\tau}(\vec{r})}\delta\rho_{\tau}(\vec{r}) + \frac{\partial\mathcal{H}^{\rm grad}(\vec{r})}{\partial\nabla\rho_{\tau}(\vec{r})}\delta\nabla\rho_{\tau}(\vec{r}) + \frac{\partial\mathcal{H}^{\rm grad}(\vec{r})}{\partial\nabla^2\rho_{\tau}(\vec{r})}\delta\nabla^2\rho_{\tau}(\vec{r}) + \cdots\bigg]d^3r\notag\\
    &~=\sum_{\tau}^{n,p}\int\sum_{n = 0}(-1)^n\nabla^n\frac{\partial\mathcal{H}^{\rm grad}(\vec{r})}{\partial\nabla^n\rho_{\tau}(\vec{r})}\delta\rho_{\tau}(\vec{r})d^3r,
\end{align}
where have integrated by parts, in order to obtain the second line.
The spatial derivative of $\rho_{\tau,\alpha}$ in Eq.~(\ref{E:pi}) is related to the spatial derivative of $S$ through
\begin{equation}
    \frac{\partial\rho_{\tau,\alpha}}{\partial\vec{r}_i} = \frac{\partial}{\partial\vec{r}_i}\sum_{\vec{r}_j\in V_{\alpha}}^{\tau_j=\tau}S(\vec{r}_j-\vec{r}_{\alpha})
     = \begin{cases}
         & \frac{\partial S(\vec{r}_i-\vec{r}_{\alpha})}{\partial\vec{r}_i},\quad \tau_i = \tau,\\
         & 0,\quad \tau_i \ne \tau.
        \end{cases}
\end{equation}
Substituting $f_{\tau}(\vec{r},\vec{p})$ expressed in Eq.~(\ref{E:f}) into Eq.~(\ref{E:HMD}), we obtain the momentum-dependent parts of the equation of motion for the test nucleons, and these are expressed in terms of the sums over the test nucleons by
\begin{align}
    \frac{\partial\mathcal{H}^{\rm MD}(\vec{r}_{\alpha})}{\partial\vec{r}_i} = &~2\frac{\partial S\big[\vec{r}_i(t) - \vec{r}_{\alpha}\big]}{\partial\vec{r}_i}\times\notag\\
    &~\bigg\{\sum_{j\in V_{\alpha}}S\big[\vec{r}_j(t) - \vec{r}_{\alpha}\big]{\mathcal{K}}_{\rm s}\big[\vec{p}_i(t),\vec{p}_j(t)\big] + \sum_{j\in V_{\alpha}}^{\tau_j = \tau_i}S\big[\vec{r}_j(t) - \vec{r}_{\alpha}\big]{\mathcal{K}}_{\rm v}\big[\vec{p}_i(t),\vec{p}_j(t)\big]\bigg\}\label{E:mdr},\\
    \frac{\partial\mathcal{H}^{\rm MD}(\vec{r}_{\alpha})}{\partial\vec{p}_i} = &~2S\big[\vec{r}_i(t) - \vec{r}_{\alpha}\big]\times\notag\\
    &~\bigg\{\sum_{j\in V_{\alpha}}S\big[\vec{r}_j(t) - \vec{r}_{\alpha}\big]\frac{\partial{\mathcal{K}}_{\rm s}\big[\vec{p}_i(t),\vec{p}_j(t)\big]}{\partial\vec{p}_i} + \sum_{j\in V_{\alpha}}^{\tau_j = \tau_i}S\big[\vec{r}_j(t) - \vec{r}_{\alpha}\big]\frac{\partial{\mathcal{K}}_{\rm v}\big[\vec{p}_i(t),\vec{p}_j(t)\big]}{\partial\vec{p}_i}\bigg\}\label{E:mdp}.
\end{align}
Based on Eqs.~(\ref{E:ri}) - (\ref{E:mdp}), one can evaluate the time evolution of coordinates and momenta $\vec{r}_i(t)$ and $\vec{p}_i(t)$ of the test nucleons, then obtain $f(\vec{r},\vec{p},t)$ through Eq.~(\ref{E:f}), based on which physical observables can be calculated.

The choice of the form factor $S(\vec{r}_i - \vec{r})$ should ensure particle number conservation
\begin{equation}
    \sum_{\alpha}\rho(\vec{r}_{\alpha})l_xl_yl_z = \frac{1}{N_{\rm E}}\sum_{\alpha}\sum_{i}S(\vec{r}_i - \vec{r}_{\alpha})l_xl_yl_z = A.
\end{equation}
In the present LBUU framework, we use a triangular form
\begin{equation}
    S(\vec{r}_i - \vec{r}) = \frac{1}{(nl/2)^6}g(\Delta x)g(\Delta y)g(\Delta z),\qquad g(q) = \Big(\frac{nl}{2} - |q|\Big)\theta\Big(\frac{nl}{2} - |q|\Big),
\end{equation}
where $\theta$ is the Heaviside function, and $n$ is an integer which determines the range of $S$.
Generally speaking, calculations on lattices violate momentum conservation since they break Galilean invariance.
Early studies have shown that the total momentum can be conserved to a high degree of accuracy if $n$ $\geqslant$ $4$~\cite{LenPRC39}.

It should be mentioned that compared with the conventional test particle method, in which the equations of motion for the test nucleons are derived from {\it single-particle} Hamiltonian, the equations of motion for the test nucleons in the LH method, Eqs.~(\ref{E:ri}) and (\ref{E:pi}), are derived from the {\it total} Hamiltonian of the system.
For the former way it is difficult to conserve energy exactly~\cite{BerPR160,LenPRC39}, while the latter can ensure the exact energy conservation , in the dynamic process~\cite{LenPRC39}.

\subsection{Collision integral}

In the present LBUU method, instead of the commonly used geometric method, the stochastic collision method~\cite{DanNPA533} is implemented for the NN collision term in the BUU equation.
In the stochastic collision approach, the collision probability of two test nucleons can be derived directly from the NN collision term, $I_{\rm c}$ in Eq.~(\ref{E:Ic}), as follows.
Considering nucleons around lattice site $\vec{r}_{\alpha}$ from two momentum space volume elements $V_{\vec{p}_1}$ $=$ $\vec{p}_1\pm\frac{1}{2}\Delta^3\vec{p}_1$ and $V_{\vec{p}_2}$ $=$ $\vec{p}_2\pm\frac{1}{2}\Delta^3\vec{p}_2$, respectively, one can average over momentum space volume $V_{\vec{p}_i}$ to obtain the distribution function $f(\vec{r}_{\alpha},\vec{p}_i)$ according to Eq.~(\ref{E:f}),
\begin{equation}\label{E:fi}
    f(\vec{r}_{\alpha},\vec{p}_i) \approx \frac{1}{\Delta^3\vec{p}_i}\frac{(2\pi\hbar)^3}{gN_{\rm E}}\sum_j^{\vec{p}_j\in V_{\vec{p}_i}}S(\vec{r}_j - \vec{r}_{\alpha}).
\end{equation}
The number of collisions between nucleons from these two momentum space volumes that happen in a time interval $\Delta t$ is
\begin{equation}\label{E:DN}
        \Delta N^{\rm coll}(\vec{r}_{\alpha}, \vec{p}_1,\vec{p}_2) = g\frac{\Delta^3\vec{p}_1}{(2\pi\hbar)^3}\Big|\frac{df(\vec{r}_{\alpha},\vec{p}_1)}{dt}\Big|_{\vec{p}_2}^{\rm coll}l_xl_yl_z\Delta t = g\frac{\Delta^3\vec{p}_2}{(2\pi\hbar)^3}\Big|\frac{df(\vec{r}_{\alpha},\vec{p}_2)}{dt}\Big|_{\vec{p}_1}^{\rm coll}l_xl_yl_z\Delta t.
\end{equation}
The quantities $\big|\frac{df(\vec{r}_{\alpha},\vec{p}_1)}{dt}\big|_{\vec{p}_2}^{\rm coll}$ and $\big|\frac{df(\vec{r}_{\alpha},\vec{p}_2)}{dt}\big|_{\vec{p}_1}^{\rm coll}$ are the changing rates of $f(\vec{r}_{\alpha},\vec{p}_1)$ and $f(\vec{r}_{\alpha},\vec{p}_2)$, respectively, caused by two-body scatterings between the nucleons in $V_{\vec{p}_1}$ and $V_{\vec{p}_2}$.
These terms can be obtained directly from Eq.~(\ref{E:Ic}), i.e., the NN collision term in the BUU equation as
\begin{align}
    \Big|\frac{df(\vec{r}_{\alpha},\vec{p}_1)}{dt}\Big|_{\vec{p}_2}^{\rm coll} =& ~g\frac{\Delta^3\vec{p}_2}{(2\pi\hbar)^3}f(\vec{r}_{\alpha},\vec{p}_1)f(\vec{r}_{\alpha},\vec{p}_2)\int\frac{d^3p_3}{(2\pi\hbar)^3}\frac{d^3p_4}{(2\pi\hbar)^3}|{\mathcal{M}}_{12\rightarrow34}|^2(2\pi)^4\delta^4(p_1 + p_2 - p_3 - p_4)\notag\\
    =&~g\frac{\Delta^3\vec{p}_2}{(2\pi\hbar)^3}f(\vec{r}_{\alpha},\vec{p}_1)f(\vec{r}_{\alpha},\vec{p}_2)v_{\rm rel}\sigma_{\rm NN}^*\label{E:df},
\end{align}
where we have substituted by the definition of the cross section
\begin{equation}
   \sigma_{\rm NN}^* = \frac{1}{v_{\rm rel}}\int\frac{d^3p_3}{(2\pi\hbar)^3}\frac{d^3p_4}{(2\pi\hbar)^3}|{\mathcal{M}}_{12\rightarrow34}|^2(2\pi)^4\delta^4(p_1 + p_2 - p_3 - p_4),
\end{equation}
with $v_{\rm rel}$ being the relative velocity of the test nucleons in the two momentum space volumes and $\sigma_{\rm NN}^*$ is the scattering cross section in the {\it two-nuclei} center-of-mass frame.
Here, we obtain the in-medium NN cross section $\sigma_{\rm NN}^*$ through multiplying the free NN cross-section $\sigma_{\rm NN}^{\rm free}$ by a medium-correction factor.
The NN elastic scattering cross section in free space $\sigma_{\rm NN}^{\rm free}$ is taken from the parameterization in Ref.~\cite{CugNIMB111}, with cutoff of $\sigma_{\rm NN}^{\rm free}(p_{\rm lab} \le 0.1~{\rm GeV}/c) = \sigma_{\rm NN}^{\rm free}(p_{\rm lab} = 0.1~{\rm GeV}/c)$ for neutron-neutron~($nn$) or proton-proton~($pp$) scatterings and
$\sigma_{\rm NN}^{\rm free}(p_{\rm lab} \le 0.05~{\rm GeV}/c) = \sigma_{\rm NN}^{\rm free}(p_{\rm lab} = 0.05~{\rm GeV}/c)$ for neutron-proton~($np$) scatterings, respectively,
since the parameterization is shown to be valid for nucleon momentum $p_{\rm lab}$ down to the corresponding cutoff~\cite{CugNIMB111}. We note that the $p_{lab}$ cutoff actually corresponds to only a few MeV of incident kinetic energy (i.e., $1.3$~MeV for $p_{\rm lab} = 0.05~{\rm GeV}/c$ and $5.3$~MeV for $p_{\rm lab} = 0.1~{\rm GeV}/c$) and these very low energy scatterings are not important in the present transport model calculations.
Since this parameterization of $\sigma_{\rm NN}^*$ is given in the {\it two-nucleon} center-of-mass frame, its value in the {\it two-nuclei} center-of-mass frame can be obtained through the Lorentz invariant quantity $E_1E_2v_{\rm rel}\sigma_{\rm NN}^*$.
From Eqs.~(\ref{E:fi})-(\ref{E:df}), one obtains
\begin{equation}
 \Delta N^{\rm coll}(\vec{r}_{\alpha}, \vec{p}_1,\vec{p}_2)
 = \sum_{i,j}^{\substack{\vec{p}_i\in V_{\vec{p}_1}\\{\vec{p}_j\in V_{\vec{p}_2}}}}\Delta N_{ij}^{\rm coll}
 = \sum_{i,j}^{\substack{\vec{p}_i\in V_{\vec{p}_1}\\{\vec{p}_j\in V_{\vec{p}_2}}}}\frac{1}{N_{\rm E}^2}v_{\rm rel}\sigma_{\rm NN}^*S(\vec{r}_i-\vec{r}_{\alpha})S(\vec{r}_j-\vec{r}_{\alpha})l_xl_yl_Z\Delta t,
\end{equation}
where the $\Delta N_{ij}^{\rm coll}$ denotes the number of physical collisions from the scattering of the $i$-th and $j$-th test nucleon.
Given that every test nucleon is $1/N_{\rm E}$ of a physical nucleon, one obtains the collision probability of the $i$-th and $j$-th test nucleons as
\begin{equation}
    P_{ij} = \frac{\Delta N_{ij}^{\rm coll}}{(1/N_{\rm E})^2} = v_{\rm rel}\sigma_{\rm NN}^*S(\vec{r}_i - \vec{r}_{\alpha})S(\vec{r}_j - \vec{r}_{\alpha})l_xl_yl_z\Delta t.
\end{equation}

One can reduce statistical fluctuations of the collision events by allowing the collisions of test nucleons that come from different ensembles.
Under such a circumstance, the collision probability is reduced, $P_{ij}$ $\rightarrow$ $P_{ij}/N_{\rm E}$ by the scaling $\sigma_{\rm NN}^*$ $\rightarrow$ $\sigma_{\rm NN}^*/N_{\rm E}$.
In our case, the NN scattering probabilities are very small within one time step, therefore instead of evaluating probabilities of all possible collisions of test nucleons, we divide randomly the test nucleons that are available for scattering around the lattice site $\alpha$ into many pairs for scattering, and accordingly amplify the corresponding scattering probabilities, which is a common scheme when allowing the scattering of test nucleons from different ensembles~\cite{DanNPA533,XZPRC71}.
The amplified scattering probabilities are given by
\begin{equation}
    P_{ij}' = P_{ij}\frac{N_{\alpha}(N_{\alpha} - 1)/2}{N_{\alpha}'/2},
\end{equation}
where $N_{\alpha}$ is the number of test nucleons that contribute to lattice site $\vec{r}_\alpha$, and $N_{\alpha}'$ is the number of test nucleons available for scattering.
Since we choose a finite range form factor for coordinate in the LBUU framework, one test nucleon can be involved in different collision events from different lattice sites.
Those test nucleons that have already collided on other lattice site are excluded in the scattering of the present lattice site, therefore $N_{\alpha}'$ is not necessarily equal to $N_{\alpha}$.
The time step $\Delta t$ needs to be sufficiently small to pin down the effect of such an exclusion by suppressing the chance of multi-scattering attempts, as well as to keep $P'_{ij}$ less than unity.
In the present LBUU framework, we choose $\Delta t$ $=$ $0.2~{\rm fm}/c$ for the full LBUU calculations, and $0.4~{\rm fm}/c$ for the Vlasov calculations (i.e., the LBUU calculations without the NN scatterings).

To verify the accuracy of the stochastic collision treatment within the present LBUU framework, we simulate collisions of nucleons confined in a cubic box of volume $V$ $=$ $10\times10\times10~\mathrm{fm}^{3}$ with periodic boundary condition.
In this simulation, we ignore the nuclear mean-field potential and the quantum nature of nucleons.
Initially, $80$ neutrons and $80$ protons are uniformly distributed over the box, corresponding to the nucleon density of $\rho=0.16~\mathrm{fm}^{3}$.
Their momenta are generated according to the relativistic Boltzmann distribution,
\begin{eqnarray}
P(p) \propto p^2\mathrm{exp}\left[ \frac{\sqrt{m^2+p^2}}{T} \right],
\end{eqnarray}
with $m$ $=$ $939~{\rm MeV}$ the free nucleon mass.
Here, the temperature $T$ is taken to be $14.24~{\rm MeV}$ so that the system has the same kinetic energy density as the zero-temperature isospin-symmetric Fermi gas of nucleons.

\begin{figure}[!htb]
\centering
\includegraphics[width=8.0cm]{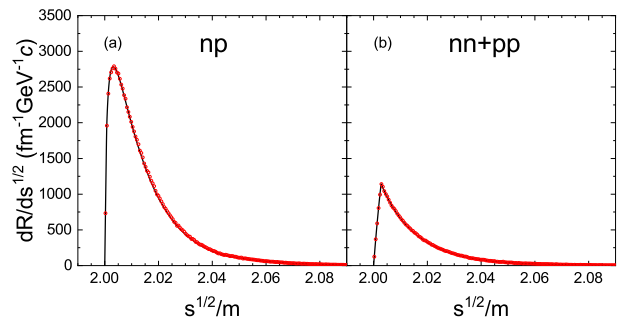}
\caption{\small The $\sqrt{s}$ distribution of (a) $np$, and (b) $nn + pp$ elastic collision rates for Boltzmann distribution at $T$ $=$ $14.24~\rm MeV$, with $80$ neutrons and $80$ protons in a cube of volume $V$ $=$ $10\times10\times10~{\rm fm}^{3}$ for theoretical~(solid line) and LBUU calculations~(circles).
In both calculations, we choose free NN cross sections as parameterized in Ref.~\cite{CugNIMB111}.}
\label{F:rt}
\end{figure}

Using the NN elastic scattering cross section in free space~\cite{CugNIMB111}, we simulate the time evolution of such a system up to $1~{\rm fm}/c$ with the time-step of $0.2~{\rm fm}/c$ and $N_{\rm{E}} = 1000$. It is constructive to see the collision rate as a function of the c.m. energy $\sqrt{s}$ of the colliding nucleon pair.
The $\sqrt{s}$ distributions of the collision rates for $np$ and $nn$ plus $pp$ are shown as red circles in the left and right panels of Fig.~\ref{F:rt}, respectively.
Theoretically, considering two species of particles following relativistic Boltzmann distributions, the $\sqrt{s}$ distributions of their collision rate can be derived as
\begin{equation}\label{E:rt}
    \frac{{\rm d}N_{\rm coll}}{{\rm d}t{\rm d}s^{1/2}} = \frac{1}{1+\delta_{ij}}\frac{N_iN_j}{V}\frac{s(s - 4m^2)K_1(s^{1/2}/T)\sigma(s^{1/2})}{4m^4T_{\rm B}K_2^2(m/T)},
\end{equation}
where $K_n$ is the $n$-th order modified Bessel function, $N_i$($N_j$) is the number of particle $i(j)$ in the volume $V$, and $\sigma$ is their scattering cross section.
The expected distributions are shown as the black solid lines in Fig.~\ref{F:rt} for comparison.
It is seen that the LBUU calculations are in excellent agreement with the expected results.

Given the quantum nature of nucleons, we treat Pauli blocking in the LBUU method as follows.
If the NN scattering between the $i$-th and $j$-th test nucleons happens on the lattice site $\vec{r}_{\alpha}$ according to $P_{ij}$ or $P'_{ij}$, the directions of their final momenta $\vec{p}_3$ and $\vec{p}_4$ are determined according to the differential cross-section given in Ref.~\cite{CugNIMB111}, and then the Pauli blocking factor $[1 - f(\vec{r}_{\alpha},\vec{p}_3)]\times[1 - f(\vec{r}_{\alpha},\vec{p}_4)]$ is used to determine whether the collision is blocked by the Pauli principle.
The distribution function $f_{\tau}(\vec{r}_{\alpha},\vec{p})$ is calculated according to Eq.~(\ref{E:fi}).
For the momentum-space volume $\Delta^3\vec{p}_i$, we adopt a sphere with radius $R_{\tau}^p(\vec{r}_{\alpha},\vec{p})$ centered at $\vec{p}_i$.
In typical BUU transport models, $R_{\tau}^p(\vec{r}_{\alpha},\vec{p})$ is a constant of about a hundred MeV.
For the calculation of small-amplitude nuclear collective dynamics near the ground state, a specifically proposed $R_{\tau}^p(\vec{r}_{\alpha},\vec{p})$ is more suitable~\cite{GaiPRC81}, i.e.,
\begin{equation}
    R_{\tau}^p(\vec{r}_{\alpha},\vec{p}) = {\rm max}[\Delta p, p_{\tau}^F(\vec{r}_{\alpha}) - |\vec{p}|],
\end{equation}
where $\Delta p$ is a constant that should be sufficiently small, and $p_{\tau}^F=\hbar(3\pi^2\rho_{\tau})^{1/3}$ is the nucleon Fermi momentum.

\subsection{Ground state initialization and evolution stability}

In the present LBUU method, we obtain the ground state of nuclei at zero temperature through varying the Hamiltonian with respect to the nuclear radial density, which is sometimes called as Thomas-Fermi~(TF) initialization~\cite{LenPRC39,DanNPA673,GaiPRC81,LHPRC99} in one-body transport model.
We assume that for a ground state nucleus at zero temperature, its Wigner function satisfies
\begin{equation}\label{E:f0}
    f_{\tau}(\vec{r},\vec{p}) = \frac{2}{(2\pi\hbar)^3}\theta\big[|\vec{p}| - p^F_{\tau}(\vec{r})\big],
\end{equation}
where $p^F_{\tau}(\vec{r})$ is local Fermi momentum given by
\begin{equation}\label{E:pF}
    p^F_{\tau}(\vec{r}) = \hbar\big[3\pi^2\rho_{\tau}(\vec{r})\big]^{1/3}.
\end{equation}
It should be noted that in principle, with the inclusion of NN scatterings, which goes beyond mean-field correlations, the nucleon momentum distribution in the ground state may differ slightly from the zero-temperature Fermi distribution.
If we assume that the nucleus is spherical for simplicity, the total energy of a ground state nucleus at zero temperature can be regarded as a functional of radial density $\rho_{\tau}(r)$ and its spatial gradients,
\begin{equation}
    E = \int{\mathcal{H}}\big[r,\rho_{\tau}(r),\nabla{\rho_{\tau}(r)},\nabla^2{\rho_{\tau}(r)}\cdots\big]dr.
\end{equation}
We can obtain the neutron~(proton) radial density in a ground state nucleus by varying the total energy with respect to $\rho_{\tau}(r)$~[note that for protons the contribution from the Coulomb interaction in Eq.~(\ref{E:Cou}) should also be included in the Hamiltonian density] as,
\begin{equation}\label{E:GS}
    \frac{1}{2m}\big\{p_{\tau}^F\big[\rho_{\tau}(r)\big]\big\}^2 + U_{\tau}\big\{p_{\tau}^{\rm F}\big[\rho_{\tau}(r)\big],r\big\} = \mu_{\tau},
\end{equation}
where $\mu_{\tau}$ is the chemical potential of proton or neutron inside the nucleus and its value is determined by the given proton number $Z$ or neutron number $N$.
The quantity $U_{\tau}\big\{p_{\tau}^{\rm F}\big[\rho_{\tau}(r)\big],r\big\}$ refers to the single nucleon potential of the nucleon with local Fermi momentum.
The single nucleon potential is derived by varying the Hamiltonian density in Eqs.~(\ref{E:Hkin})-(\ref{E:HDD}) with respect to the phase space distribution function and density gradients, and its detailed expression for the N$3$LO Skyrme pseudopotential is shown in Ref.~\cite{WRPRC98}.
The physical significance of Eq.~(\ref{E:GS}) is very intuitive: in a classical picture, in a ground state nucleus at zero temperature, the nucleons in the Fermi surface of different radial position have the same chemical potential.
The local density $\rho_{\tau}(\vec{r})$ for a ground state spherical nucleus is obtained by solving Eq.~(\ref{E:GS}) subject to the following boundary condition on the total local density $\rho(r)$ $=$ $\rho_n(r)$ $+$ $\rho_p(r)$,
\begin{equation}
    \frac{\partial\rho(r)}{\partial r}\Big|_{r = 0} = \frac{\partial\rho(r)}{\partial r}\Big|_{r = r_{\rm B}} = 0.
\end{equation}
Here, $r_{\rm B}$ is the boundary of the nucleus and it satisfies $\rho(r_{\rm B}) = 0$.

In the present LBUU framework, the initial coordinates of test nucleons are generated according to the obtained $\rho_{\tau}(\vec{r})$, while their initial momenta to a zero-temperature Fermi distribution with the Fermi momentum given in Eq.~(\ref{E:pF}).
Due to the presence of the form factor $S(\vec{r} - \vec{r}')$ introduced in Eq.~(\ref{E:f}), the density is smeared slightly in the LBUU calculations compared with the realistic local density.
Thus the initial ground state radial density distribution is slightly different from the solution of Eq.~(\ref{E:GS}).
Unlike the Gaussian wave packet which is used to mimic the Wigner function in QMD model~\cite{AicPR202}, the form factor $S(\vec{r} - \vec{r}')$ does not have any physical meaning, and it can be regarded as a numerical technique introduced in the test-particle approach so that one can obtain well-defined densities and mean fields.
As shown in the following, an additional gradient term in the local density can compensate for the effects caused by the smearing of the local density due to the form factor.
In the following of this subsection, we will use $\tilde{\rho}(\vec{r})$ to represent the local density in the LBUU calculation while $\rho(\vec{r})$ to the realistic local density.
The local density $\tilde{\rho}(\vec{r})$ can be regarded as a convolution of the realistic local density with the form factor,
\begin{equation}
    \tilde{\rho}(\vec{r}) = \int\rho(\vec{r}')S(\vec{r} - \vec{r}')d^3r'.
\end{equation}
To express $\rho(\vec{r})$ in terms of $\tilde{\rho}(\vec{r})$, we have formally
\begin{align}
    \rho(\vec{r}) & = \int\tilde{\rho}(\vec{r}')S^{-1}(\vec{r}' - \vec{r})d^3r' = \int\Big[\sum_{n = 0}^{\infty}\frac{1}{n!}\nabla^n\tilde{\rho}(\vec{r})(\vec{r}' - \vec{r})^n\Big]S^{-1}(\vec{r}' - \vec{r})d^3r'\notag\\
    & \approx\tilde{\rho}(\vec{r}) + c\nabla^2\tilde{\rho}(\vec{r})\label{E:rho},
\end{align}
where we have truncated at next-to-leading order~[the $\nabla\tilde{\rho}(\vec{r})$ term vanishes because of the symmetry of the integral], and $S^{-1}(\vec{r} - \vec{r}')$ is the inverse of $S(\vec{r} - \vec{r}')$ which satisfies
\begin{equation}
    \int S(\vec{r} - \vec{r}'')S^{-1}(\vec{r}'' - \vec{r}')d^3r'' = \delta(\vec{r} - \vec{r}').
\end{equation}
The parameter $c$ defined by
\begin{equation}
    c \equiv \int\frac{1}{2}(\vec{r}' - \vec{r})^2S^{-1}(\vec{r}' - \vec{r})d^3r',
\end{equation}
is a small constant that only depends on the form of $S$.
In the LBUU framework, to obtain $\rho(\vec{r})$ by the direct correction on $\tilde{\rho}(\vec{r})$, is not feasible since numerically the density in Eq.~(\ref{E:rho}) is not always positive.
If we substitute Eq.~(\ref{E:rho}) into the total Hamiltonian, with several necessary approximations, we obtain an additional term that is proportional to $c\tilde{\rho}(\vec{r})\nabla^2\tilde{\rho}(\vec{r})$.
This term leads to an additional gradient term $\tilde{E}^{[2]}\nabla^2\tilde{\rho}$ in the equations of motion Eq.~(\ref{E:pi}).
Therefore, in practice we can add the additional gradient terms $\tilde{E}^{[2]}\nabla^2\tilde{\rho}$ to the equations of motion, to compensate for the smearing of density due to the form factor.
In principle, the parameter $\tilde{E}^{[2]}$ should contain higher order effects, therefore we adjust it to roughly obtain the ground state rms radius evolution with the smallest oscillation, since the rms radius in the exact ground state should not change with time.
Normally $\tilde{E}^{[2]}$ is a small parameter around $15~\rm MeV$ for various (N$3$LO)~Skyrme parameter sets.
It should be mentioned that this correction on the density gradient term only improves the stability of the ground state evolution~(rms radius and radial density profile) slightly, and does not cause much difference on the results for collective motions.
In ideal cases with $N_{\rm E}$ $\rightarrow$ $\infty$ and $l_x,l_y,l_z$ $\rightarrow$ $0$, the local density in LBUU calculation will approach the physical local density, and $\tilde{E}^{[2]}$ will become zero.
Since all the LBUU calculations are based on $\tilde{\rho}(\vec{r})$, we do not distinguish $\tilde{\rho}(\vec{r})$ and $\rho(\vec{r})$, and $\rho(\vec{r})$ should be interpreted as $\tilde{\rho}(\vec{r})$ in the rest of the article.

\begin{figure}[!htb]
\centering
\includegraphics[width=8.0cm]{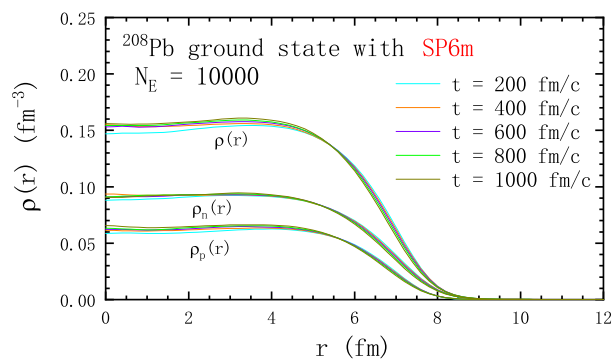}
\caption{Time evolution of the radial density profile of the ground state of \isotope[208]{Pb} based on the LHV calculation (i.e., the LBUU calculation without NN scatterings) with N$3$LO Skyrme pseudopotential SP$6$m up to $1000~{\rm fm}/c$.
Taken from Ref.~\cite{WRPRC99} with permission from the American Physical Society.}
\label{F:DP}
\end{figure}

We first examine the ground-state evolution stability of the LHV calculation, i.e., the LBUU calculation without the collision term, since in principle all NN scatterings should be blocked in the ground state.
We show in Fig.~\ref{F:DP} the time evolution of the radial density profile from the LHV calculation for the nucleus \isotope[208]{Pb} in ground state up to $1000~{\rm fm}/c$, obtained with $N_{\rm E}$ $=$ $10000$ and a time step $\Delta t$ $=$ $0.4~{\rm fm}/c$ by using the N$3$LO Skyrme pseudopotential SP$6$m.
We notice from Fig.~\ref{F:DP} that the profile of the radial density exhibits only very small variations with time, which indicates the success of the above initialization method.
It also shows that the smearing of the local density caused by the inclusion of the form factor $S$ does not affect the dynamic evolution significantly.
Such features indicate that the present LBUU method of solving the BUU equation can be used to study long-time nuclear processes such as nuclear spallation and heavy-ion fusion reactions.

\begin{figure}[!htp]
\centering
\includegraphics[width=12.0cm]{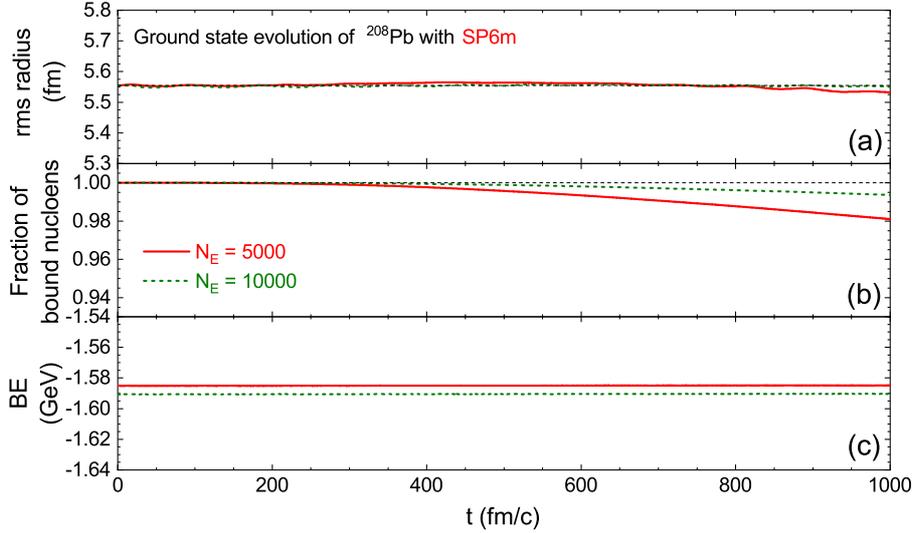}
\caption{Time evolution of (a)~rms radius, (b) fraction of bound nucleons and (c) binding energy of \isotope[208]{Pb} ground state from the LHV calculation (i.e., the LBUU calculation without NN scatterings) with the N$3$LO Skyrme pseudopotential SP$6$m up to $1000~{\rm fm}/c$.
Calculations are performed with a time step $\Delta t$ $=$ $0.4~{\rm fm}/c$, and $N_{\rm E}$ $=$ $5000$ and $10000$, respectively.
Taken from Ref.~\cite{WRPRC99} with permission from the American Physical Society.}
\label{F:TE}
\end{figure}

Apart from the radial density profile, other properties concerning the ground-state evolution stability are also examined.
In Fig.~\ref{F:TE}, we present the time evolution of the rms radius, the fraction of bound nucleons, and the binding energy of the the LHV calculation (i.e., the LBUU calculation without NN scatterings).
The calculations are performed with time step $\Delta t$ $=$ $0.4~{\rm fm}/c$, and $N_{\rm E}$ $=$ $5000$ and $10000$, respectively.
The test nucleons whose form factor do not overlap with that of others are considered as free test nucleons, and they are excluded in calculating the fraction of bound nucleons and the rms radius.
We notice from Fig.~\ref{F:TE}(a) that although in $N_{\rm E} = 5000$ case, the rms radius starts to decrease after about $800~{\rm fm}/c$, the LHV calculation gives a fairly stable time evolution of rms radius.
This decrease is due to the evaporation of test nucleons from the bound nuclei, which is illustrated in Fig.~\ref{F:TE}(b).
Such an evaporation of test nucleons is inevitable in transport model calculations due to the limited precision in the numerical realization, whereas it can be suppressed by increasing $N_{\rm E}$, as can be seen in Fig.~\ref{F:TE}(b), though the result with $E_{\rm E}$ $=$ $5000$ is already satisfactory\cite{WRPRC99}.
As shown in Fig.~\ref{F:TE}(c), the LH method ensures the energy conservation to a very high degree.
The difference between the cases $N_{\rm E}$ $=$ $5000$ and $N_{\rm E}$ $=$ $10000$ is mainly caused by the numerical precision of the gradient term in the Hamiltonian.
It is seen from Fig.~\ref{F:TE} that the present LBUU framework can give a fairly stable ground-state time evolution.
Due to the high efficiency of the GPU parallel computing, it becomes possible to include more ensembles or test particles in the LBUU calculation.
As one will see in the following, to obtain the correct GDR width, as much as $30000$ ensembles are needed in the full LBUU calculation with NN scatterings.

For the stability of the ground-state evolution in the full LBUU calculation, we note that for $\sigma_{\rm NN}^{\rm free}$ with $N_{\rm E} = 30000$, the rms radius and the ground-state energy of \isotope[208]{Pb} vary less than 3.6\% (0.2 fm) and 3.2\% (50 MeV), respectively, during the time evolution of $0$ $-$ $500$ ${\rm fm}/c$~\cite{WRX19}.
The stability of the rms radius in the full LBUU calculation is not as good as in the LHV case, and this may be caused by the fact that with the inclusion of NN scatterings, i.e., beyond mean-field correlations, the nucleon momentum distribution in the ground state may differ slightly from the zero-temperature Fermi distribution.
Apart from this, although the LH method can conserve the energy almost exactly for the mean-field evolution without NN collisions, the non-perfect energy conservation in the LBUU calculation might be caused by the NN scattering processes, which usually violate the energy conservation when the momentum-dependent mean-field potentials are used.
Both problems require further investigation of the transport model calculations in the future.

\subsection{Nuclear giant resonances within transport models}

We consider a small excitation of the Hamiltonian
\begin{equation}
    \hat{H}_{ex}(t) = \lambda\hat{Q}\delta(t - t_0),
\end{equation}
where $\hat{Q}$ is the excitation operator for a given mode, and $\lambda$ is the initial excitation parameter supposed to be small.
Within linear response theory~\cite{Fet1971}, the response of the excitation operator $\hat{Q}$ as a function of time is given by
\begin{equation}
    \Delta\langle\hat{Q}\rangle(t) = \langle 0'|\hat{Q}|0'\rangle(t) - \langle0|\hat{Q}|0\rangle(t) = -\frac{2\lambda\theta(t)}{\hbar}\sum_F|\langle F|\hat{Q}|0\rangle|^2{\rm sin}\frac{(E_F-E_0)t}{\hbar},
    \label{E:dQt}
\end{equation}
where $|0\rangle$ is the unperturbed nuclear ground state with energy $E_0$, $|0'\rangle$ is the nuclear state after the perturbation, and $|F\rangle$ is the energy eigenstate of the excited nucleus with eigen-energy $E_F$.
The strength function, which is defined as
\begin{equation}
    S(E) = \sum_F|\langle F|\hat{Q}|0\rangle|^2\delta(E - E_F + E_0),
\end{equation}
can be expressed as a Fourier integral of $\Delta\langle\hat{Q}\rangle(t)$ in Eq.~(\ref{E:dQt}) by
\begin{equation}\label{E:S-Q}
    S(E) = -\frac{1}{\pi\lambda}\int_0^{\infty}dt\Delta\langle\hat{Q}\rangle(t){\rm sin}\frac{Et}{\hbar}.
\end{equation}
By evaluating the time evolution of $\Delta\langle\hat{Q}\rangle(t)$ within the transport model, we can obtain the strength function, and subsequently other quantities like the peak energy, width and energy-weighted sum rules.
The time evolution of $\Delta\langle\hat{Q}\rangle(t)$ can be expressed in terms of the Wigner function $f(\vec{r},\vec{p})$ as follows.

If we assume $\hat{Q}$ is a one-body operator, then it can be written as the sum of single particle operators $\hat{q}$ acting on each nucleon, $\hat{Q}$ $=$ $\sum_i^A\hat{q}$, and the expectation value of $\hat{Q}$ for a given state is evaluated as follows,
\begin{align}
    \langle\hat{Q}\rangle = &~\langle \Phi|\hat{Q}|\Phi\rangle = \int\langle \Phi|\vec{r}_1\cdots\vec{r}_N\rangle\langle\vec{r}_1\cdots\vec{r}_N|\hat{Q}|\vec{r}_1'\cdots\vec{r}_N'\rangle\langle\vec{r}_1'\cdots\vec{r}_N'|\Phi\rangle d^3r_1\cdots d^3r_Nd^3r_1'\cdots d^3r_N'\label{E:Qt},
\end{align}
where we have added two identity operators.
Considering the definition of the one-body density matrix,
\begin{equation*}
    \rho(\vec{r}_1,\vec{r}_1') = A\int\langle\vec{r}_1\vec{r}_2\cdots\vec{r}_N|\Phi\rangle\langle\Phi|\vec{r}_1'\vec{r}_2\cdots\vec{r}_N\rangle d^3r_2\cdots d^3r_N,
\end{equation*}
and combining it with the one-body operator condition $\hat{Q}$ $=$ $\sum_i^A\hat{q}$, we can rewrite Eq.~(\ref{E:Qt}) as
\begin{equation}
    \langle\hat{Q}\rangle = \int\rho(\vec{r}_1',\vec{r}_1)\langle\vec{r}_1|\hat{q}|\vec{r}_1'\rangle d^3r_1d^3r_1'\label{E:Q2}.
\end{equation}
The density matrix can be expressed in coordinate space as the inverse Fourier transform of $f(\vec{r},\vec{p})$ by
\begin{equation}
    \rho\Big(\vec{r} - \frac{\vec{s}}{2},\vec{r} + \frac{\vec{s}}{2}\Big) = \int f(\vec{r},\vec{p}){\rm exp}\Big(i\frac{\vec{p}}{\hbar}\vec{s}\Big)d^3p\label{E:Q3}.
\end{equation}
In the above equation we have changed the integration variables, $\vec{r}_1$ $=$ $\vec{r} + \frac{\vec{s}}{2}$ and $\vec{r}_1'$ $=$ $\vec{r} - \frac{\vec{s}}{2}$.
We define the Wigner transform of $\hat{q}$ in coordinate space,
\begin{equation}
    q(\vec{r},\vec{p}) \equiv \int{\rm exp}\Big(-i\frac{\vec{p}}{\hbar}\cdot\vec{s}\Big)q\Big(\vec{r}+\frac{\vec{s}}{2},\vec{r}-\frac{\vec{s}}{2}\Big)d^3s\label{E:Q4},
\end{equation}
where $q\big(\vec{r} + \frac{\vec{s}}{2},\vec{r} - \frac{\vec{s}}{2}\big)$ $=$ $\big\langle\vec{r} + \frac{\vec{s}}{2}|\hat{q}|\vec{r} - \frac{\vec{s}}{2}\big\rangle$ represents the matrix element of $\hat{q}$ in coordinate space.
Substituting Eq.~(\ref{E:Q3}) and the inverse transform of Eq.~(\ref{E:Q4}) into Eq.~(\ref{E:Q2}), the expectation of $\hat{Q}$ can be written into the following form,
\begin{equation}
    \langle\hat{Q}\rangle = \int f(\vec{r},\vec{p})q(\vec{r},\vec{p})d^3rd^3p\label{E:Q5},
\end{equation}
which means the time evolution of $\langle\hat{Q}\rangle$ can be calculated through the time evolution of $f(\vec{r},\vec{p})$.

In the transport model, different external excitation $\lambda\hat{Q}\delta(t-t_0)$ can be generated by changing the positions and momenta of the test nucleons as follows~\cite{UrbPRC85}:
\begin{equation}\label{E:q}
    \vec{r}_i \longrightarrow \vec{r}_i + \lambda\frac{\partial q(\vec{r}_i,\vec{p}_i)}{\partial\vec{p}_i},\quad\quad \vec{p}_i \longrightarrow \vec{p}_i - \lambda\frac{\partial q(\vec{r}_i,\vec{p}_i)}{\partial\vec{r}_i}.
\end{equation}
The detailed forms of $q(\vec{r}_i,\vec{p}_i)$ for different collective modes and their corresponding initialization in the transport model will be given later.

\section{Lattice Hamiltonian Vlasov calculations}\label{S:LHV}

In this section we compare the peak energy of nuclear giant resonances obtained from LBUU calculations without the NN scatterings, i.e., the LHV calculations, with that from the RPA, since the $2p$-$2h$ correlation is absent in both cases.
Both the isoscalar monopole and the isovector dipole modes of \isotope[208]{Pb} are examined.

\subsection{\label{S:MS}Isoscalar monopole mode}

Since the isoscalar giant monopole resonance~(ISGMR) provides information about the nuclear matter incompressibility~\cite{YouPRL82,ShlPRC47,LTPRL99,PatPLB718,PatPLB726,GupPLB760}, which is a fundamental quantity that characterizing the EOS of symmetric nuclear matter, it is interesting to study the ISGMR within the transport model to make a cross check with the incompressibility extracted from the HICs.

From point of view of the one-body transport model, the isoscalar monopole mode is regarded as a compressional breathing of the nuclear fluid.
The excitation operator for the isoscalar monopole mode $\hat{Q}_{\rm ISM}$ and its one-body operator $\hat{q}_{\rm ISM}$ take the form
\begin{equation}
    \hat{Q}_{\rm ISM} = \frac{1}{A}\sum_i^{\rm A}\hat{r}_i^2\label{E:QMS},\quad\hat{q}_{\rm ISM} = \frac{\hat{r}^2}{A}.
\end{equation}
Through Eq.~(\ref{E:Q4}), we obtain the Wigner transform of $\hat{q}_{\rm ISM}$ as
\begin{equation}\label{E:qISM}
    q_{\rm ISM}(\vec{r},\vec{p}) = \frac{\vec{r}^2}{A}.
\end{equation}
According to Eq.~(\ref{E:q}), we can generate in the transport model the initial isoscalar monopole excitation by changing the initial phase space information of test nucleons with respect to that of the ground state,
\begin{equation}
    \vec{p}_i\longrightarrow\vec{p}_i - 2\lambda\frac{\vec{r}_i}{A}.
\end{equation}
The spatial coordinates of the test nucleons remain unchanged since $q_{\rm ISM}$ in Eq.~(\ref{E:qISM}) is independent of momentum.
Note that the rms radius of a nucleus, shown in Fig.~\ref{F:TE}, is given by the square root of the expectation value of $\hat{Q}_{\rm ISM}$.

\begin{figure}[!h]
\centering
\includegraphics[width=10.0cm]{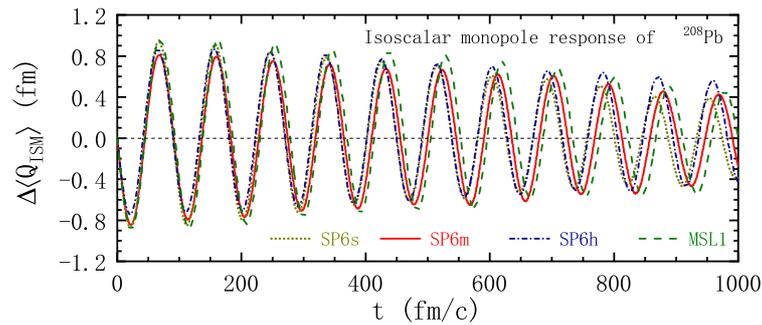}
\caption{Time evolution of $\Delta\langle\hat{Q}_{\rm ISM}\rangle$ of \isotope[208]{Pb} after a perturbation by $\hat{H}_{ex}(t)$ $=$ $\lambda\hat{Q}_{\rm ISM}\delta(t-t_0)$ with $\lambda$ $=$ $100~{\rm MeV\cdot  fm^{-1}}/c$ in the LHV calculations.
The results correspond to three N$3$LO Skyrme pseudo potentials, SP$6$s, SP$6$m, SP$6$h, and one conventional Skyrme interaction MSL$1$, respectively.
Taken from Ref.~\cite{WRPRC99} with permission from the American Physical Society.}
\label{F:QMS}
\end{figure}

We show in Fig.~\ref{F:QMS} the time evolution of $\Delta\langle\hat{Q}_{\rm ISM}\rangle$, i.e., the difference of the expectation value of $\langle\hat{Q}_{\rm ISM}\rangle$ between the excited and the ground state from the LHV calculations.
The results are from one conventional Skyrme interaction MSL$1$, and three N$3$LO Skyrme pseudopotentials, SP$6$s, SP$6$m, and SP$6$h.
In the calculation, we set the number of ensembles $N_{\rm E}$ to be $5000$, and the initial excitation parameter $\lambda$ to be $100~{\rm MeV\cdot fm^{-1}}/c$.
One sees from the figure that the time evolution of $\Delta\langle \hat{Q}_{\rm ISM}\rangle$, or equivalently the rms radius, displays a very regular oscillation, and the quick increase of the radius with time, which is generally seen in most BUU calculations using the conventional test particle method, does not show up here.
Besides that, since the only damping mechanism in the LHV calculation is Landau damping, the amplitude of the oscillation only decreases slightly.
Landau damping is caused by one-body dissipation which is governed by a coupling of single-particle and collective motion.
It should be mentioned that in the RPA framework, the damping also comes only from one-body dissipation, since the coupling to more complex states, like $2p$-$2h$ states, is missing in RPA~\cite{BerRMP55}.
We obtain the peak energy of the GMR through Fourier transform of the time evolution of $\Delta\langle\hat{Q}_{\rm ISM}\rangle$ shown in Fig.~\ref{F:QMS}.
The obtained peak energy is $13.8~\rm MeV$ for SP$6$s, $13.6~\rm MeV$ for SP$6$m, $13.9~\rm MeV$ for SP$6$h, and $13.5~\rm MeV$ for MSL$1$.
In order to compare the result from the LHV calculation with that from RPA, we calculate the strength function of GMR using the Skyrme-RPA code by Colo {\it et al.}~\cite{ColCPC184} with the MSL1 interaction.
The obtained peak energy $14.1~\rm MeV$ is comparable to that from the LHV calculation with MSL$1$, and the small discrepancy may reflect the difference between the semi-classical and quantum nature.

\subsection{\label{S:DV}Isovector dipole mode}

The isovector giant dipole resonance~(IVGDR) of finite nuclei is the earliest observed nuclear collective excitation.
Systematic experimental investigation of the IVGDR with photon-nuclear reactions has been done decades ago~\cite{BerRMP47}.
Recent precise measurements of the isovector dipole response have been performed at RCNP for \isotope[48]{Ca}~\cite{BirPRL118}, \isotope[120]{Sn}~\cite{HasPRC92}, and \isotope[208]{Pb}~\cite{TamPRL107} with inelastic proton scattering, as well as at GSI for \isotope[68]{Ni}~\cite{RosPRL111} by using Coulomb excitation in inverse kinematics.
Recently, a low-lying mode called pygmy dipole resonance~(PDR) has been observed experimentally~\cite{RyePRL89,AdrPRL95,WiePRL102,EndPRL105}, and this effect has already been studied based on the Vlasov equation~\cite{UrbPRC85}.
The IVGDR\cite{YilPRC72,TriPRC77,KHYPRC95}, PDR~\cite{CarPRC81,BarPRC88}, and electric dipole polarizability $\alpha_D$~\cite{PiePRC85,RocPRC88,RocPRC92,ZZPRC92}, which are dominated by these isovector dipole modes, provide sensitive probes to constrain the density dependence of the nuclear symmetry energy.

\begin{figure}[!h]
\centering
\includegraphics[width=10.0cm]{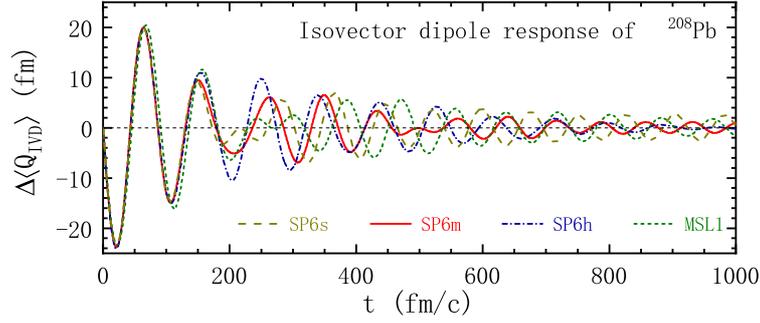}
\caption{Same as Fig.~\ref{F:QMS} but for isovector dipole mode with $\lambda$ $=$ $25~{\rm MeV}/c$.
Taken from Ref.~\cite{WRPRC99} with permission from the American Physical Society.}
\label{F:QDVLHV}
\end{figure}

For the isovector dipole mode, the external perturbation can be written in the following form,
\begin{equation}
    \hat{Q}_{\rm IVD} = \frac{N}{A}\sum_i^{\rm Z}\hat{z}_i - \frac{Z}{A}\sum_i^{\rm N}\hat{z}_i\label{E:QDV}.
\end{equation}
The coefficients in front of the single particle position operator are defined to keep the center of mass of the nucleus stays at rest.
According to Eq.~(\ref{E:q}), the excited nucleus can be obtained in transport models by changing the initial phase space coordinates of test nucleons,
\begin{equation}
    p_z \longrightarrow
    \begin{cases}
         &p_z - \lambda\frac{N}{A},\quad \rm for~protons,\\
         &p_z + \lambda\frac{Z}{A},\quad \rm for~neutrons.
    \end{cases}
\end{equation}
We show in Fig.~\ref{F:QDVLHV} the time evolution of $\Delta\langle\hat{Q}_{\rm IVD}\rangle$ for \isotope[208]{Pb} with the interactions SP$6$s, SP$6$m, SP$6$h, and MSL$1$ of the LHV calculations.
The number of ensembles $N_{\rm E}$ and the initial excitation parameter $\lambda$ are set to be $5000$ and $25~{\rm MeV}/c$, respectively.
Based on the time evolution of $\Delta\langle\hat{Q}_{\rm IVD}\rangle$ shown in Fig.~\ref{F:QDVLHV}, the obtained peak energy for SP$6$s, SP$6$m, SP$6$h and MSL$1$ are $13.4~\rm MeV$, $13.5~\rm MeV$, $13.7~\rm MeV$ and $13.1~\rm MeV$, respectively.
The peak energies of MSL$1$ from the RPA calculation is $13.3~\rm MeV$, which is comparable to that from the LBUU calculation without the NN collision term.

\section{Spreading width of giant dipole resonance and collisional damping}\label{S:width}

It is generally thought that in low-energy HICs with incident energy of only a few MeV/nucleon, the NN scattering can be safely neglected since they are mostly blocked by the Pauli principle.
However, when it comes to the width of the GDR , the collisional damping caused by NN scatterings is an essential mechanism to enhance the insufficient GDR width obtained through the pure Vlasov calculation~\cite{WRX19}.
Nevertheless, to properly implement the damping mechanism caused by NN scatterings in transport models requires a rather accurate treatment of the Pauli blocking, which is a challenge to transport model calculations.
The difficulty mainly comes from calculating local momentum distributions $f_{\tau}(\vec{r}_{\alpha},\vec{p})$ accurately in transport models.
The inaccuracy of $f_{\tau}(\vec{r}_{\alpha},\vec{p})$ affects the accuracy of the Pauli blocking and leads to spurious collisions, which enhance the collisional damping and thus overestimate the width of nuclear giant resonances.
There are three main origins of the inaccuracy of the calculated $f_{\tau}(\vec{r}_{\alpha},\vec{p})$ and thus the spurious collisions in transport models:

1) Fluctuations in calculating $f_{\tau}(\vec{r}_{\alpha},\vec{p})$ through Eq.~(\ref{E:fi}) caused by too small $N_{\rm E}$;

2) Spurious temperature caused by a finite $\Delta p$ in calculating $f_{\tau}(\vec{r}_{\alpha},\vec{p})$~(also see Ref.~\cite{GaiPRC81});

3) The finite lattice spacing $l$ causes the diffusion in local momentum space due to the average of different local lattice densities in the nuclear surface region.

In order to obtain the spreading width with high accuracy with the BUU equation, one should choose a large $N_{\rm{E}}$ together with the sufficiently small $l$ and $\Delta p$.
After a careful test, it is found~\cite{WRX19} that to get a convergent GDR width, $l$ should be smaller than $0.5~\rm fm$, $\Delta p$ smaller than $0.05~\rm GeV$, and $N_{\rm E}$ larger than $30000$.
Further reducing $\Delta p$ and $l$ or increasing $N_{\rm E}$ leads only to a negligible decrease of the calculated GDR width.
Therefore, in the following full LBUU calculations of the GDR width, we take
$l=0.5~\rm fm$, $\Delta p=0.05~\rm GeV$, and $N_{\rm E}=30000$.

\begin{figure}[!hbt]
\centering
\includegraphics[width=12.0cm]{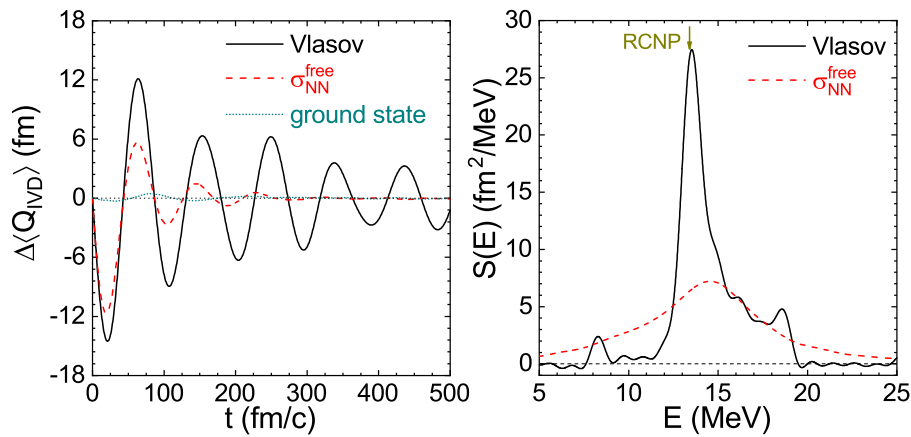}
\caption{Time evolution of $\Delta\langle\hat{Q}_{\rm IVD}\rangle$~(left) and its strength function~(right) of \isotope[208]{Pb} after a perturbation $\hat{H}_{ex}$ $=$ $\lambda\hat{Q}_{\rm IVD}\delta(t - t_0)$ with $\lambda$ $=$ $15{\rm MeV}/c$ from the LHV (Vlasov) calculation and the LBUU calculation with the free NN cross section $\sigma_{\rm NN}^{\rm free}$. The dotted line in the left window is the expectation value of $\hat{Q}_{\rm IVD}$ in the ground state from the LBUU calculation with $\sigma_{\rm NN}^{\rm free}$.
Taken from Ref.~\cite{WRX19} under the Creative Commons CCBY license.}
\label{F:QDV}
\end{figure}

The collisional damping or NN scattering can have a significant effect on the width of nuclear giant resonances.
Shown in Fig.~\ref{F:QDV} is the time evolution of isovector dipole response $\Delta\langle\hat{Q}_{\rm IVD}\rangle$ of \isotope[208]{Pb} and its strength function from the LHV calculation and the full LBUU calculation with the free NN elastic scattering cross section~\cite{WRX19}.
In both cases, the N$3$LO Skyrme pseudopotential SP$6$h is adopted, and the same initial excitation with $\lambda$ $=$ $15~{\rm MeV}/c$ is employed~(we note varying $\lambda$ by $2/3$ leads to almost the same value of the GDR width).
Shown in the left window of Fig.~\ref{F:QDV} by the dotted line is the time evolution of the expectation value $\langle0|\hat{Q}_{\rm IVD}|0\rangle$ in the ground state of \isotope[208]{Pb} as obtained from the LBUU calculation with the free-space NN cross section.
The expectation value $\langle0|\hat{Q}_{\rm IVD}|0\rangle$ in the ground state of \isotope[208]{Pb} is negligible compared with that in the GDR cases with and without NN scatterings.
It is seen that including NN scatterings enhances the damping of the oscillations significantly , and leads to a much larger width.
From the strength functions, the obtained GDR width of $^{208}$Pb is $1.5~\rm MeV$ in the Vlasov calculation and $6.5~\rm MeV$ in LBUU calculation with NN scatterings.
We also notice from the right panel of Fig.~\ref{F:QDV} that the peak shifts to higher energy when we include the NN scatterings.
The impact of NN scatterings on the width indicates that they might also affect some particular observables in low-energy HICs, e.g., the nuclear stopping of HICs in Fermi energy region, which can be studied further.

\begin{figure}[!hbt]
\centering
\includegraphics[width=9.0cm]{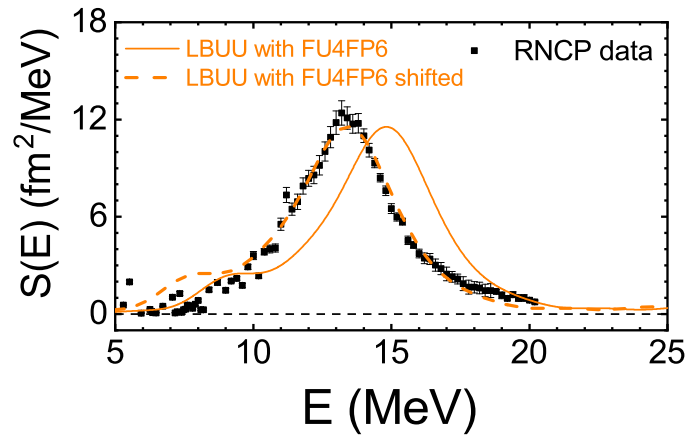}
\caption{Strength function of the GDR in \isotope[208]{Pb} after a perturbation $\hat{H}_{ex}$ $=$ $\lambda\hat{Q}_{\rm IVD}\delta(t - t_0)$ with $\lambda$ $=$ $15{\rm MeV}/c$ from the LBUU calculation using the FU$4$FP$6$ parameterization~\cite{OLCPC43} for the in-medium NN scattering cross section.
The strength function measured in the RCNP experiment~\cite{TamPRL107} is also shown for comparison.}
\label{F:SE}
\end{figure}

Recent experiments of the $\isotope[208]{Pb}(\vec{p},\vec{p}')$ reaction performed at RCNP~\cite{TamPRL107} have measured the GDR width of \isotope[208]{Pb} accurately, and a value of $4.0~\rm MeV$ has been extracted.
Therefore, the LBUU calculation with the free NN elastic cross section (which predicts a GDR width of $6.5~\rm MeV$) significantly overestimates the GDR width of \isotope[208]{Pb}.
It is well known that the NN elastic cross section is suppressed in the nuclear medium, therefore the overestimation of the GDR width with $\sigma_{\rm NN}^{\rm free}$ is understandable since the medium effects on the NN elastic cross section will weaken the collisional damping and consequently result in smaller GDR width.
As shown in Ref.~\cite{WRX19}, in order to reproduce the experimental GDR width of \isotope[208]{Pb} of RCNP, a strong medium reduction of the NN cross section is needed.
There are many parameterizations for the medium reduction of the NN cross section~\cite{LopPRC90,LPPRC97,BarPRC99,OLCPC43}, which could be density-, collision energy-, isospin-dependent.
As an example, we choose the FU$4$FP$6$ parameterization~\cite{OLCPC43} for the medium reduction of the cross section to calculate the strength function and width of the GDR in \isotope[208]{Pb}.The FU$4$FP$6$ parameterization of the medium reduction is density-, momentum- and isospin-dependent, and it is preferred by the nucleon induced nuclear reaction cross section data~\cite{OLCPC43} and predicts a very strong in-medium reduction of NN scattering cross sections.
The strength function of the GDR in \isotope[208]{Pb} from the LBUU calculation is shown in Fig.~\ref{F:SE} and compared with the RCNP data~\cite{TamPRL107}.
The obtained GDR width from the LBUU calculation through the full width at half maximum~(FWHM) of the strength function is $4.32~\rm MeV$, which is consistent with $4.0~\rm MeV$ from the RCNP experiment.
However, the Skyrme pseudopotential SP$6$h adopted in the calculation overestimates the peak energy by about $1.5~\rm MeV$.
Using effective interactions with a different symmetry energy slope parameter $L$ or nuclear effective masses can easily reproduce the correct peak energy~\cite{KHYPRC95}.
In order to compare the shape of the strength function and the value of the width of the GDR in \isotope[208]{Pb}, we shift the strength function from LBUU calculation to meet the experimental peak energy.
We conclude from Fig.~\ref{F:SE} that the present LBUU method with the FU$4$FP$6$ parameterization~\cite{OLCPC43} for the medium reduction of the NN scattering cross section can nicely reproduce the measured shape of the strength function and the width of GDR in \isotope[208]{Pb}.
It should be stressed that the FU$4$FP$6$ parameterization suggests a very strong in-medium reduction of NN scattering cross sections, consistent with the conclusions obtained in Ref.~\cite{WRX19}.

\section{Summary and outlook}\label{S:S&O}

We have reviewed the recent progress in calculating the nuclear collective motions by solving the BUU equation with the LH method.
In order to calculate the nuclear collective motions accurately with the BUU equation, the present LBUU framework contains the following features:
1) the smearing of the local density is considered in the equations of motion self-consistently through the lattice Hamiltonian method;
2) the initialization of a ground state nucleus is carried out according to a nucleon radial density distribution obtained by varying the same Hamiltonian that governs the evolution;
3) the NN collision term in the BUU equation is implemented through a full-ensemble stochastic collision approach;
4) the high-performance GPU parallel computing is employed to increase the computing efficiency.
The present LBUU framework with these features brings us to a new level of precision in solving the BUU equation.

Within the LBUU framework, it has been shown that the peak energies of ISGMR and IVGDR from the pure Vlasov calculation are consistent with the RPA calculation, and the full LBUU calculation is able to give reasonable GDR strength function compared with the experimental data.
The peak energies can be used to extract information about the nuclear EOS, while the width of the GDR can constrain the medium reduction of the elastic NN scattering cross section.
The success of the present LBUU framework in describing the nuclear collective motions has demonstrated its capability in treating the stability of ground-state nuclei and the nuclear dynamics near equilibrium.
Therefore the present LBUU framework provides a solid foundation for studying the long-time process of heavy-ion reactions at low energies, e.g., heavy-ion fusion and multi-nucleon transfer reactions at near-barrier energies, based on solving the BUU equation.
The significant effects of the collisional damping on the width of the nuclear GDR indicate that NN scatterings should play a crucial role in nuclear collective dynamics with small amplitude oscillations.

The present LBUU framework has been shown to significantly reduce the uncertainties of the transport model simulations for HICs in various aspects, especially for the stability of the nuclear ground state evolution and the very accurate treatment of NN scatterings as well as the Pauli blocking.
This is rather important for various studies of HICs based on transport model calculations, e.g., the extraction of the nuclear EOS and the in-medium NN scattering cross sections.
Further studies of HICs from low to intermediate energies within the present LBUU framework are in progress and it is expected that more reliable information on the nuclear EOS, the in-medium NN scattering cross sections, and the effective nuclear interactions could be extracted in near future.

\section*{Conflict of Interest Statement}

The authors declare that the research was conducted in the absence of any commercial or financial relationships that could be construed as a potential conflict of interest.



\section*{Funding}
This work was partially supported by the National Natural Science Foundation of China under Contracts No. $11947214$, No. 11905302, No. $11890714$, No. $11625521$ and No. $11421505$, the Key Research Program of Frontier Sciences of the CAS under Grant No. QYZDJ-SSW-SLH$002$, the Strategic Priority Research Program of the CAS under Grants No. XDB$16$ and No. XDB$34000000$, and the Major State Basic Research Development Program (973 Program) in China under Contract No. $2015$CB$856904$.

\section*{Acknowledgments}
We thank Pawel Danielewicz, Che Ming Ko, Bao-An Li and Jun Su for helpful discussions,
and Meisen Gao, Jie Pu, Xiaopeng Zhang, Chen Zhong and Ying Zhou for setting up and maintaining the GPU severs.

\section*{Data Availability Statement}
The data sets for this study are available from the corresponding author upon reasonable request.

\bibliographystyle{frontiersinHLTH&FPHY} 





\end{document}